\documentclass[11pt,a4paper]{article}
\usepackage{mathrsfs}
\usepackage{dutchcal}
\usepackage{amsfonts}
\usepackage[hscale=0.8,vscale=0.8]{geometry}
\newcommand{\B}{\scriptscriptstyle B}

\begin{document}

\date{}

\title{Correction to: The Double-Wedge Algebra for Quantum Fields
on Schwarzschild and Minkowski Spacetimes}
\author{Bernard S.~Kay \\ 
\normalsize Department of Mathematics, 
University of York, York YO10 5DD, U.K.} 

\maketitle

\abstract{\footnotesize There is an error in the proof (but not the truth) of Theorem 3.2 in the author's 1985 paper ``The Double-Wedge Algebra for Quantum Fields on Schwarzschild and Minkowski Spacetimes'' in ``Communications in Mathematical Physics''.  The author became aware of that error and of how it may be corrected soon after it went to print, and two companion papers published soon afterwards (in Helvetica Physica Acta) refer to an ``erratum to appear''.    This is that erratum.   We also take the opportunity to note a few (unrelated and minor) corrections to those two companion papers and also to very briefly mention related more recent work.}

\section*{}

There is an error in the proof (but not the truth) of Theorem 3.2 in \cite{double}.  The error is on the 5th line of the proof where it is written:

\medskip

\noindent
``because $h_{\B}$ and hence $\tilde h_{\B}$ have no zero eigenvalues, $d\Gamma(\tilde h_{\B})\psi=\psi \Rightarrow \psi = \Omega$.''

\medskip

We should say straight away that this does not even correctly express what the author had had in mind at the time.  Rather the author's intended meaning at the time of writing would have been correctly expressed, instead, by

\medskip

\noindent
``because $h_{\B}$ and hence $\tilde h_{\B}$ have no zero eigenvalues, $d\Gamma(\tilde h_{\B})\psi=0 \Rightarrow \psi = \lambda\Omega$.''

\medskip

\noindent
where (it should have been said both in Theorem 1.3 and in the proof of Theorem 3.2) $\lambda$
is some complex number.  

But, even after making this correction, the above replacement statement is false in general.  To see that it is false, suppose that, on the one-particle Hilbert space $\mathcal{h}$, $h_{\B}$ had an eigenvector, $x$, with a non-zero eigenvalue, say $a$ (but does not have zero as an eigenvalue).

Then consider the two-particle state $\psi = (x \oplus 0)\otimes (0 \oplus x)$ in the two-particle Hilbert subspace of the Fock space over $\tilde{\mathcal{h}}$. Clearly we would have
\[
d\Gamma(\tilde h_{\B})\psi= (h_{\B} x \oplus 0) \otimes (0 \oplus x) + (x \oplus 0) \otimes (0 \oplus -h_{\B} x) = a (x \oplus 0) \otimes (0 \oplus x) - a (x \oplus 0) \otimes (0 \oplus x) = 0.
\]

However, the proof of the theorem can be salvaged by invoking the fact that $h_{\B}$ actually has no eigenvectors at all which easily follows from the last equality in Appendix A3 of \cite{double}.

The proof of Theorem 3.2 may thus be fixed by replacing the above quote by:

\medskip

\noindent
``because $h_{\B}$ and hence $\tilde h_{\B}$ have no eigenvectors at all, $d\Gamma(\tilde h_{\B})\psi=0 \Rightarrow \psi = \lambda\Omega$ (where $\lambda$ is some complex number).''

\medskip

\noindent
which is easily seen to be a valid statement.

Actually,  as is explained and proved  in the companion paper \cite{pure} (see Note (6) on page 1034 and Note (2) on page 1037 of that paper) in the ``quasi-free Bose case'' (in the language of that paper) which is the only case relevant to \cite{double}, the results of Theorem 1.3 of \cite{double}, and hence an essentially more elementary alternative proof of Theorem 3.2 of \cite{double} can be had by invoking Theorem 2 (on page 1037) of \cite{pure} and this allows one to dispense with Condition (a) of Theorem 1.3 of \cite{double} (= Condition ($\alpha$) of Theorem 1 in \cite{pure}) -- i.e.\ to dispense with the need to show that $d\Gamma(\tilde h_{\B})\psi=0 \Rightarrow \psi = \lambda\Omega$.  In this alternative proof, one in fact just needs that the one particle Hamiltonian, $h_{\B}$ (and hence $\tilde h_{\B}$) has no zero eigenvalue, which is, as may be seen from \cite{pure} (or from the other companion paper \cite{unique}) part of the definition of ``ground one-particle structure''.

\medskip

\noindent
The need for the above (twofold) correction to the 5th line of the proof of Theorem 3.2 of \cite{double} was actually noticed by the author soon after \cite{double} went to the printer and before the companion papers \cite{unique} and \cite{pure} were finalized and, indeed, both those papers cite \cite{double} together with an ``erratum to appear''.   While I have privately make the content of that erratum known to a number of people over the years, I neglected to publish it until now.

\medskip

\noindent
I take the opportunity to also mention here a few (much more minor) errors in \cite{pure} and \cite{unique}: 

\smallskip

\noindent
In \cite{pure}: There is a tilde missing over a Gothic $\mathfrak{A}$ on the last line of page 1031. Also the name ``Winnink'' is misspelled in reference [6].  

\smallskip

\noindent
In \cite{unique}: In Theorem 1a, ``given dynamical system $(D, \mathscr{T}(t))$'' should be ``given dynamical system $(D, \sigma, \mathscr{T}(t))$''.  There is a tilde missing on the second occurrence of the symbol $\mathscr{H}$ in Definition 1b.  Also, in the same definition, ``$K^\beta: D\rightarrow \tilde H$'' should be ``$K^\beta: D\rightarrow \tilde\mathscr{H}$''.   ``Lemma 5.1'' on page 1022  should be ``Lemma 4.1''.  

\medskip

\noindent
Lastly, we very briefly comment on the relation between \cite{double} (and \cite{pure} and \cite{unique}) and subsequent work.   (See also e.g.\ \cite{KayEncyc2ndEd} for further discussion.)  The results in \cite{double} obviously beg the question of how the Hartle-Hawking-Israel state constructed there can be extended from the double wedge to the whole of Kruskal.   This problem was solved by Sanders \cite{SandersHHI} and, using different techniques, by G\'erard \cite{GerardHHI}.  (As briefly discussed \ in \cite{KayEncyc2ndEd}, each of the two constructions generalize to a different class of spacetimes.)

The results and techniques in the three papers \cite{double,unique,pure}, and in particular, the notions of `double linear dynamical system' and `double KMS one-particle structure' (for linear quantum dynamical systems) and `double quantum dynamical system' and `double KMS state' (for general quantum dynamical systems) are expected to be useful in a number of other contexts, and have, for example, found application in the construction of scalar quantum fields with polynomial interactions in 2-dimensional de Sitter space -- see the monograph of Barata, J\"akel and Mund \cite{BJM}.   

As far as I am aware, none of the subsequent work which makes use of, or refers to, \cite{double} is affected by the above-corrected error.

\end{document}